\documentclass
[twocolumn,superscriptaddress,amsmath,amssymb,aps,prl,float]{revtex4-1}%
\usepackage{graphicx}
\usepackage{dcolumn}
\usepackage{bm}
\usepackage{hyperref}%
\usepackage{amsmath}%
\usepackage{amsfonts}%
\usepackage{amssymb}
\newcommand{\xmark}{$\mathsf{X}$ }

\providecommand{\U}[1]{\protect\rule{.1in}{.1in}}

\begin{document}
\title{Reduction of the spin susceptibility in the superconducting state of Sr$_{2}%
$RuO$_{4}$ observed by polarized neutron scattering}
\author{A.\,N. Petsch}
\email{a.petsch@bristol.ac.uk}
\author{M.\,Zhu}
\affiliation{H.H. Wills Physics Laboratory, University of Bristol, Bristol BS8 1TL, United Kingdom}
\author{Mechthild Enderle}
\affiliation{Institut Laue-Langevin, CS 20156, 38042 Grenoble Cedex 9, France}
\author{Z.\,Q. Mao}
\affiliation{Department of Physics, Graduate School of Science, Kyoto University, Kyoto
606-8502, Japan}
\affiliation{Department of Physics, Pennsylvania State University, University Park,
Pennsylvania 16802, USA}
\author{Y. Maeno}
\affiliation{Department of Physics, Graduate School of Science, Kyoto University, Kyoto
606-8502, Japan}
\author{I.\,I. Mazin}
\affiliation{Department of Physics and Astronomy, George Mason University
and Quantum Science and Engineering Center, Fairfax, VA
22030, USA}
\author{S.\,M. Hayden}
\email{s.hayden@bristol.ac.uk}
\affiliation{H.H. Wills Physics Laboratory, University of Bristol, Bristol BS8 1TL, United Kingdom}
\date{\today}

\begin{abstract}
Recent observations [A.~Pustogow \emph{et al}.~Nature \textbf{574}, 72 (2019)]
of a drop of the $^{17}$O nuclear magnetic resonance (NMR) Knight shift in the
superconducting state of Sr$_{2}$RuO$_{4}$ challenged the popular picture of a
chiral odd-parity paired state in this compound. Here we use polarized neutron
scattering (PNS) to show that there is a $34 \pm6$\,\% drop in the magnetic
susceptibility at the Ru site below the superconducting transition
temperature. We measure at lower fields $H \sim\tfrac{1}{3} H_{c2}$
than a previous PNS study allowing the suppression to be observed. The PNS
measurements show a smaller susceptibility suppression than NMR measurements
performed at similar field and temperature. Our results rule out the chiral
odd-parity $\mathbf{d}=\hat{\mathbf{z}}(k_{x}\pm ik_{y})$ state and are
consistent with several recent proposals for the order parameter including
even-parity $B_{1g}$ and odd-parity helical states.

\end{abstract}
\maketitle


\textit{Introduction.}---Sr$_{2}$RuO$_{4}$ is a moderately correlated oxide
metal which, which forms a good Fermi liquid and
superconducts~\cite{Maeno1994_MHYN,Maeno1994_MHYN} below 1.5\thinspace K. It
has been initially proposed as a solid state
analogue~\cite{Rice1995_RiSi,Mazin1997} of superfluid $^{3}$He--A, driven by
proximity to ferromagnetism. The superconducting state was widely assumed to
possess chiral odd-parity order~\cite{Mackenzie2003_MaYo,Maeno2012} with
broken time-reversal symmetry~\cite{Luke1998_LFKL}. An important property of
odd-parity (triplet-paired) superconductors is that for some magnetic field
directions the spin susceptibility may show no change upon entering the
superconducting state. This property may be investigated by probes not
sensitive to the superconducting diamagnetic screening currents, such as
nuclear magnetic resonance (NMR) or polarized neutron scattering (PNS). Early
studies of the susceptibility using the NMR Knight shift with $^{17}$O
(NMR)~\cite{Ishida1998_IMKA} and PNS~\cite{Duffy2000_DHMM} detected no change
while crossing the superconducting transition when magnetic fields were
applied parallel to the RuO$_{2}$-planes or $ab$-planes. These observations
supported the picture of triplet-paring with an out-of-plane $\mathbf{d}%
$-vector or an unpinned in-plane $\mathbf{d}$-vector $\perp\mathbf{H}$. A
muon-spin rotation ($\mu$SR) study~\cite{Luke1998_LFKL} observed that magnetic
moments appeared below the superconducting transition temperature $T_{c}$.
This was interpreted as evidence for time-reversal symmetry breaking (TRSB) in
the superconducting phase. Further evidence for TRSB came from the detection
of a Kerr effect~\cite{Xia2006_XMBF} associated with the superconducting transition.

More recently, it has become clear that it is difficult to consistently
describe the physical properties of the superconducting state with a simple
odd-parity representation~\cite{Mackenzie2017_MSHM}. For example, the favored
$\mathbf{d}=\hat{\mathbf{z}}(k_{x}\pm ik_{y})$ state implies the existence of
edge currents which are not detected experimentally \cite{Kirtley2007_KKHK,
Hicks2010_HKLK, Curran2011_CKBG, Curran2014_CBDG} and $H_{c2}$ is much lower
than expected~\cite{Mackenzie2017_MSHM}. Also an NMR experiment failed
to detect any changes in susceptibility for a $c$-axis field  \cite{Murukawa},
even though it would have to be reduced below $T_{c}$ in the $\hat{\mathbf{z}%
}(k_{x}\pm ik_{y})$ state. It was shown that rotation of the order parameter
vector would be forbidden by the strong spin-orbit
coupling \cite{Zutic,Kim2017}, which led one of us to conclude that
``the Knight shift in Sr$_{2}$RuO$_{4}$ remains a challenge
for theorists; until this puzzle is resolved, we cannot use the Knight shift
argument''\cite{Zutic}.

Further progress has been made recently with a $^{17}$O-NMR study by Pustogow
et al.~\cite{Pustogow2019_PLCS} which detected a significant reduction in the
Knight shift on entering the superconducting state for in-plane fields for the
first time. This result has now been reproduced by Ishida et
al.~\cite{Ishida2019_IsMM}. The observed reduction in the Knight shift on
entering the superconducting state of Sr$_{2}$RuO$_{4}$ shows that the spin
susceptibility is reduced. This should also be observed in polarized neutron
scattering measurements. However, a PNS study~\cite{Duffy2000_DHMM} with
relatively poor statistics and at a field $\mu_{0}H=1$\thinspace T was unable
to observe a reduction. 

In this paper, we report PNS measurements at a lower field ($\mu_{0}%
H=0.5$\thinspace T) and with better statistics. We find a $34\pm6$\thinspace\%
drop in the magnetic susceptibility at the Ru site below the superconducting
transition temperature. This is somewhat smaller than the $63\pm8$\thinspace\%
drop observed by NMR Knight shift measurements \cite{Ishida2019_IsMM} at the
in-plane oxygen site for a similar field $\mu_{0}H=0.48$\thinspace T. We discuss possible reasons for the difference between the two
observations and the constraints our results place on the allowed order
superconducting order parameter.

\begin{table*}[t]
\begin{ruledtabular}
\begin{tabular}{c@{}c@{}c@{}c@{}c@{\ \ \ \ }c@{\ \ \ \ }c||c@{}c@{}c@{}c@{}c@{\ \ \ \ }c@{\ \ \ \ }c}
state & basis & line & TRSB & \multicolumn{3}{c||}{$\chi_0(H,T\rightarrow0)/\chi_0(n)$} & state & basis & line & TRSB & \multicolumn{3}{c}{$\chi_0(H,T\rightarrow0)/\chi_0(n)$} \\
 & function & nodes & & \multicolumn{3}{c||}{with $\mathbf{H}\!\parallel$} & & function & nodes & & \multicolumn{3}{c}{with $\mathbf{H}\!\parallel$} \\
 & $\Delta(\mathbf{k})$ & & & $[100]$ & $[110]$ & $[001]$  & & $\mathbf{d}(\mathbf{k})$ & & & $[100]$ & $[110]$ & $[001]$\\ \hline
 $A_{1g}$  & $k_x^2+k_y^2$ & & \xmark & 0 & 0 & 0 & $A_{1u}$ & $\hat{\mathbf{x}}k_x+\hat{\mathbf{y}}k_y$ & & \xmark & 1/2 & 1/2 & 1 \\
 & & & & & &  & & $\hat{\mathbf{z}}k_z$ & (h) & \xmark & 1 & 1 & 0 \\
 $A_{2g}$ & $k_xk_y(k_x^2-k_y^2)$ & (v) & \xmark & 0 & 0 & 0 & $A_{2u}$ & $\hat{\mathbf{x}}k_y-\hat{\mathbf{y}}k_x$ & & \xmark & 1/2 & 1/2 & 1 \\
 & & & & & &  & & $\hat{\mathbf{z}}k_xk_yk_z(k_x^2-k_y^2)$ & (v,h) & \xmark & 1 & 1 & 0 \\
 $B_{1g}$ & $k_x^2-k_y^2$ & (v) & \xmark & 0 & 0 & 0 & $B_{1u}$ & $\hat{\mathbf{x}}k_x-\hat{\mathbf{y}}k_y$ &  & \xmark & 1/2 & 1/2 & 1 \\
& & & & & &  & & $\hat{\mathbf{z}}k_xk_yk_z$ & (v,h) & \xmark & 1 & 1 & 0 \\
 $B_{2g}$ & $k_xk_y$ & (v) & \xmark & 0 & 0 & 0 & $B_{2u}$ & $\hat{\mathbf{x}}k_y+\hat{\mathbf{y}}k_x$ &  & \xmark & 1/2 & 1/2 & 1 \\
& & & & & &  & & $\hat{\mathbf{z}}k_z(k_x^2-k_y^2)$ & (v,h) & \xmark & 1 & 1 & 0 \\
 $E_{g}(a)$ & $k_xk_z$;\;$k_yk_z$ & (v,h) & \xmark & 0 & 0 & 0  & $E_{u}(a)$ & $\hat{\mathbf{x}}k_z$;\;$\hat{\mathbf{y}}k_z$ & (h) & \xmark & 0;\;1 & $1/2$ & 1 \\
 & & & & & &  & & $\hat{\mathbf{z}}k_x$;\;$\hat{\mathbf{z}}k_y$ & (v) & \xmark & 1 & 1 & 0 \\
 $E_{g}(b)$ & $(k_x\pm k_y)k_z$ & (v,h) & \xmark & 0 & 0 & 0  & $E_{u}(b)$ & $(\hat{\mathbf{x}}\pm\hat{\mathbf{y}})k_z$ & (h) & \xmark & $1/2$ & 0;\;1 & 1 \\
 & & & & & &  & & $\hat{\mathbf{z}}(k_x\pm k_y)$ & (v) & \xmark & 1 & 1 & 0 \\
 $E_{g}(c)$ & $(k_x\pm ik_y)k_z$ & (h) & \checkmark & 0 & 0 & 0 & $E_{u}(c)$ & $(\hat{\mathbf{x}}\pm i\hat{\mathbf{y}})k_z$ & (h) & \checkmark & 1/2 & 1/2 & 1 \\
 & & & & & &  & & $\hat{\mathbf{z}}(k_x\pm ik_y)$ & & \checkmark & 1 & 1 & 0
\end{tabular}
\end{ruledtabular}
\caption{Irreducible representations of superconducting order parameters
compatible with the tetragonal point group $D_{4h}$ with strong spin-orbit
coupling~\cite{Annett1990_Anne}. The left and right panels are even-parity
(singlet) and odd-parity (triplet) states, respectively.  Column
three and ten show whether states have vertical line nodes, $\| k_{z}$, (v),
or horizontal line nodes, $\bot k_{z}$ (h) on a 2D Fermi surface. States with
$k_{\mu}$ transform like $\sin k_{\mu}$ while states with $k_{\mu}^{2}$
transform like $\cos k_{\mu}$. Ticks indicate whether TRSB is present.
$\chi_{0}$ is calculated from Eqn.~\ref{Eqn:leggett} in the $H \rightarrow0, T
\rightarrow0$ limit, for a cylindrical Fermi surface.}%
\label{tab:irr_reps}%
\end{table*}

\textit{The spin susceptibility as a probe of the superconducting
state.}---The transition to a superconducting state involves the pairing of
electrons and hence a change to the spin wavefunction. In the case of singlet
pairing where the spin susceptibility is suppressed everywhere (barring exotic
cases as Ising superconductivity) its temperature dependence is described by
the Yosida function~\cite{Yosida1958_Yosi,Won1994_WoMa}. A triplet
superconductor is described by a spinor order parameter,
\begin{equation}
\mathbf{\Delta}%
=\begin{bmatrix} \Delta_{\uparrow\uparrow} & \Delta_{\uparrow\downarrow} \\ \Delta_{\downarrow\uparrow} & \Delta_{\downarrow\downarrow} \end{bmatrix}=\Delta
\begin{bmatrix} -d_x+id_y & d_z \\ d_z & d_x+id_y \end{bmatrix},
\end{equation}
where the $\mathbf{d}$-vector is $\mathbf{d}=(d_{x},d_{y},d_{z})$. The
non-interacting spin susceptibility tensor is given by
\cite{Leggett1975_Legg}
\begin{eqnarray}
\label{Eqn:leggett} 
\!\frac{\chi_{0,\alpha\beta}}{\chi_{0}(n)}\!=\!\delta_{\alpha\beta}%
+\!\int\!\frac{d\Omega}{4\pi}\,[Y(\hat{\mathbf{k}},T)\!-\!1]\;\Re\!\left\{
\!\frac{d_{\alpha}^{\ast}(\mathbf{k})d_{\beta}(\mathbf{k})}{\mathbf{d}^{\ast
}\!(\mathbf{k})\!\cdot\!\mathbf{d}(\mathbf{k})}\!\right\}  \!,
\end{eqnarray}
where the integral is over the Fermi surface, $Y(\hat{\mathbf{k}},T)$ is the
Yosida function and $\chi_{0}(n)$ the normal-state spin susceptibility.
Table\ \ref{tab:irr_reps} shows $\chi_{0}(T\rightarrow0,H\rightarrow
0)=\chi_{0}(0)$ evaluated using Eqn.~\ref{Eqn:leggett} for selected
irreducible representations and applied field directions.

The measurements of the spin susceptibility in the superconducting state are
complicated by diamagnetic screening due to supercurrents which precludes
quantitative measurements with standard techniques such as SQUID magnetometry.
The NMR Knight shift and PNS have been used successfully to measure the spin
susceptibility in the superconducting state. The Knight shift, $K$, originates
from the hyperfine interaction between the nuclear moment and the magnetic
field at the nuclear site produced by the electrons surrounding that site,
which is only indirectly related to the total magnetization. For instance,
core polarization and spin-dipole interaction do not contribute to the latter,
but affect the Knight shift, while spin and orbital moments have different
spatial distributions and therefore produce different contributions to $K$
with opposite signs in some cases~\cite{Pavarini2006,Yongkang2019}. In
contrast, PNS probes the total magnetization density $M(\mathbf{r})$ in
absolute units induced by an external magnetic field $\mu_{0}H$. The orbital
and spin magnetization are equally weighted and $M(\mathbf{r})$ is averaged in
space. In the normal state of Sr$_{2}$RuO$_{4}$, three bands cross the Fermi
surface~\cite{Oguchi1995}. The partially filled Ru $t_{2g}$ orbitals account
for the majority of the density of states at the Fermi energy. Hence the
majority of the spin susceptibility is associated with the Ru site probed by PNS.

In PNS, the spatially varying density $M(\mathbf{r})$ is measured by
diffraction. This technique was first applied to V$_{3}$Si by Shull and
Wedgwood~\cite{Shull1966_ShWe}. It has also been used to probe
cuprate~\cite{Boucherle1993a} and iron-based~\cite{Lester2011_LCAS,
Brand2014_BSWH} superconductors where singlet pairing has been observed. Early
PNS measurements by~\citet{Shull1966_ShWe} of the temperature dependence of
the induced magnetization in V$_{3}$Si showed that the susceptibility only
dropped to about $\tfrac{1}{3}$ of its normal state value for applied fields
of $\approx0.1H_{c2}$. This residual susceptibility is due to the orbital
susceptibility present in transition metals~\cite{Clogston1962a,White1983}.
Importantly, such cancellation effects work very differently in NMR and PNS,
for instance, in V metal NMR shows no or a very small change \cite{Noer1964,Garifullin2008} in the Knight shift across
$T_{c},$ because of nearly exact cancellation of the core polarization (an
effect specific to NMR) and the Fermi-contact term.

NMR and PNS probe susceptibility of superconducting in the mixed state and
typically in relatively high magnetic fields $\sim1$\ T. It is well known that
vortices create low-energy electronic states ~\cite{Caroli1964a,Volovik1993a}.
In the mixed state of conventional superconductors the associated
quasiparticle density of states $\mathcal{N}_{F}^{\star}(H)\propto
\mathcal{N}_{F}^{\star}(n)H/H_{c2}$ comes from low-energy localized states in
the vortex cores~\cite{Caroli1964a}, while in superconductors with lines of
gap nodes $\mathcal{N}_{F}^{\star}(H)\propto\mathcal{N}_{F}^{\star}%
(n)\sqrt{H/H_{c2}}$ comes from the vicinity of the gap nodes in the momentum
space, and partially from outside the vortex cores~\cite{Volovik1993a}. The
same states give rise to a linear heat capacity and also contribute to the
spin susceptibility. For example, a linear field dependence of $\chi=M/H$ has
been observed \cite{Lester2011_LCAS} in the superconducting state of
Ba(Fe$_{1-x}$Co$_{x}$)$_{2}$As$_{2}$.

\begin{figure*}[t]
\centering
\includegraphics[width=0.9\linewidth]{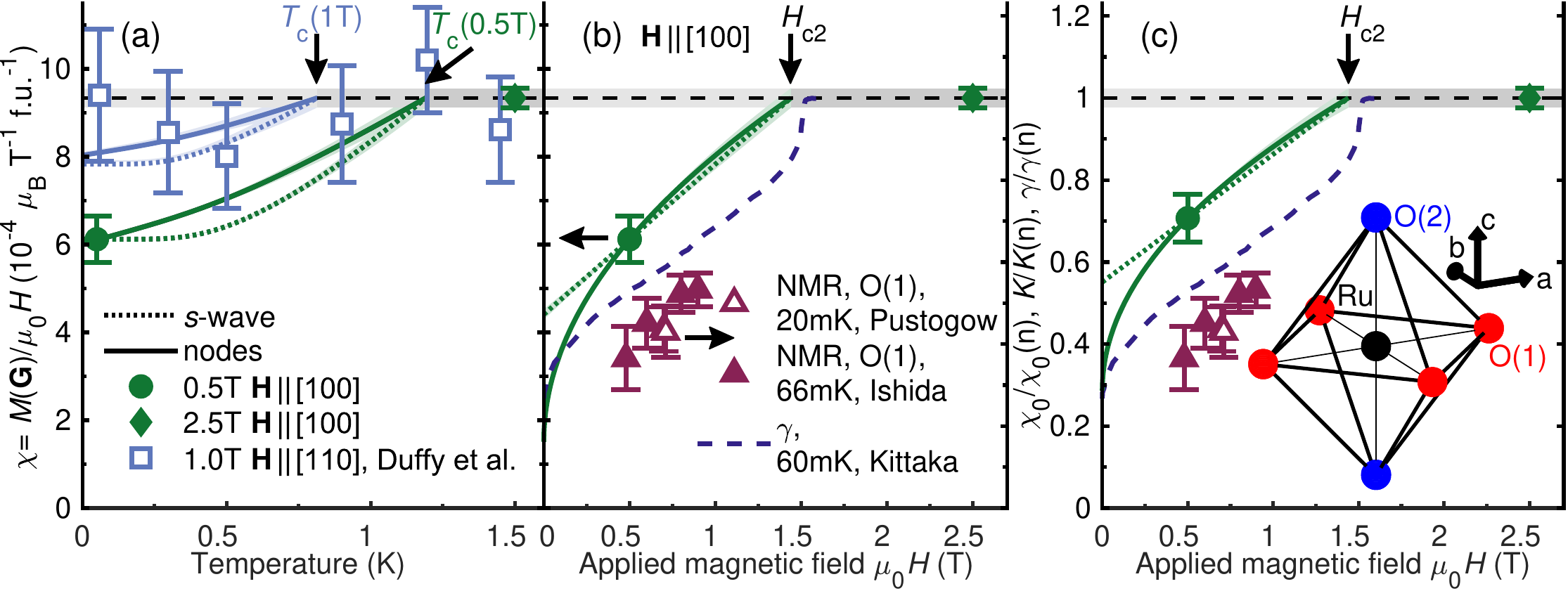}\vspace{10pt}
\vspace{0pt}\caption{PNS measurements of the susceptibility $\chi
=M(\mathbf{G})/\mu_{0} H$ at $\mathbf{G}=(011)$. (a) $T$-dependence of $\chi$
shows a drop below $T_{c}$ for $\mu_{0}\mathbf{H}\parallel[100]$. Results of
Duffy et al.~\cite{Duffy2000_DHMM} with $\mu_{0}\mathbf{H}=1\text{T}%
\parallel[110]$ (open symbols) are shown scaled by the Ru magnetic form
factor~\cite{Brown1992_Brow}. $T$-dependencies (dotted and solid lines) are
calculated using the Yosida functions
\cite{Yosida1958_Yosi,Won1994_WoMa,Anderson1961_AnMo} and $\chi(H)$ from (b).
(b) $H$-dependence of $\chi$ at $T=60$\,mK. $\chi(H)$ is fitted to simple
$s$-wave and nodal models (see text). (c) $H$-dependence of the scaled
non-interacting spin susceptibility $\chi_{0}(H)/\chi_{0}(n)$ determined from
(b) by correction for Stoner enhancement and orbital contribution
\cite{Supp_mat}. Also shown are $^{17}$O NMR Knight shift, $K$, data at
similar temperatures from Pustogow et al.\cite{Pustogow2019_PLCS} and Ishida
et al.\cite{Ishida2019_IsMM}, and the measured linear coefficient of the
specific heat, $\gamma=C_{e}/T$ from Kittaka et al.\cite{Kittaka2018_KNSK}.
Both quantities have been scaled to their normal state values.}%
\label{fig:Figure}%
\end{figure*}

\textit{Experimental method.}---The experimental set up and crystal were the
same as described in Duffy et al.~\cite{Duffy2000_DHMM}. A single crystal
(C117) of Sr$_{2}$RuO$_{4}$ with dimensions of 1.5\thinspace mm$\times
$2\thinspace mm$\times$5\thinspace mm was cut from a crystal grown by
floating-zone method and glued with GE-varnish on a copper stage with [100]
aligned vertically. The sample size was chosen so as not to saturate the
detector. To ensure good thermal contact on the sample the copper stage was
connected via two 1~mm diameter copper wires to the dilution refrigerator.
$T_{c}$ of this sample is 1.47\thinspace K and $\mu_{0}H_{c2}($100\thinspace
mK$)=1.43$\thinspace T for $\mathbf{H}||[110]$%
~\cite{Duffy2000_DHMM,Mao2000_MaMF}. We used the three-axis spectrometer IN20
at the Institut Laue-Langevin, Grenoble~\cite{HAYDEN2019_HEPM}. A vertical
magnetic field was applied perpendicular to the scattering plane along the
[100] direction. Measurements of the nuclear Bragg reflections (002) and (011)
verified that the field was within $0.11^\circ$ of the [100] direction. The PNS experiments were performed in the
superconducting state at the (011) Bragg reflection with $\mu_{0}%
H=0.5$\thinspace T. The beam was monochromatic with $E=63$\thinspace meV. The
spectrometer beam polarization was 93.2$\pm$0.1\%, measured at the (011)
reflection using a Heusler monochromator and analyzer. Detector counts were normalized either to time or by using a neutron counter placed in the incident beam. Both methods produced consistent results. The sample was field cooled and test
measurements with polarization analysis were performed to check for depolarization
from the vortex lattice. We measured the beam polarization at the (002) and
(011) Bragg reflections for $\mu_{0}H=0.5$\thinspace T, no measurable
difference between superconducting and normal states could be detected.
Susceptibility measurements were performed with a polarized beam without the
analyzer as in Duffy et al.~\cite{Duffy2000_DHMM}.

Polarized neutron scattering experiments can directly probe the real-space
magnetization density $\mathbf{M}(\mathbf{r})$ in the unit cell, induced by a
large magnetic field $\mu_{0}H$. Further details of the method and theory are
given in our previous studies~\cite{Duffy2000_DHMM,Lester2011_LCAS}. Due to
the periodic crystal structure, the applied magnetic field induces a
magnetization density with spatial Fourier components $\mathbf{M}(\mathbf{G}%
)$, where $\mathbf{G}$ are the reciprocal lattice vectors, and
\begin{equation}
\mathbf{M(G)}=\int_{\mbox{unit cell}}\mathbf{M(r)}\exp(i\mathbf{G\cdot
r})\mathrm{{d}\mathbf{r}.}\label{Eq:M_G}%
\end{equation}
We measure the flipping ratio $R$, defined as the ratio of the cross-sections,
$I_{+}$, $I_{-}$, of polarized incident beams with neutrons parallel or
anti-parallel to the applied magnetic field. A detector insensitive to the
scattered spin polarization and summing over the final spin states was used. Because the induced moment is small, the experiment is
carried out in the limit $(\gamma r_{0}/2\mu_{B})M(\mathbf{G})/F_{N}%
(\mathbf{G})\ll1$. In this limit \cite{Duffy2000_DHMM}, the flipping ratio
is,
\begin{equation}
R=1-\frac{2\gamma r_{0}}{\mu_{B}}\frac{{M}(\mathbf{G})}{F_{N}(\mathbf{G}%
)},\label{Eq:Flipping}%
\end{equation}
where the nuclear structure factor $F_{N}(\mathbf{G})$ is known from the
crystal structure and $\gamma r_{0}=5.36\times10^{-15}$\thinspace m. For the
(011) reflection we used $F_{N}=4.63\times10^{-15}\,\text{m f.u.}^{-1}$.

\textit{Results.}---In the present experiment we apply magnetic fields along
the [100] direction to allow comparison with NMR measurements
\cite{Pustogow2019_PLCS,Ishida2019_IsMM}. We first established the normal
state susceptibility $\chi(n)$ by making measurements at $T=1.5$~K and
$\mu_{0}H=2.5$\thinspace T as shown in Fig.~\ref{fig:Figure}(a). Our signal,
$1-R$, is proportional to the induced moment (Eq.~\ref{Eq:Flipping}) and our
error bars are determined by the number of counts in $I_{\pm}$ and hence the
counting time. Thus, the 2.5\thinspace T measurement (closed diamond) provides
the most accurate estimate of the normal state susceptibility, shown by the
horizontal solid line. The data has been converted to $\chi$ using
Eqn.~\ref{Eq:Flipping} \cite{cor_note}. A further
measurement (closed circle) was performed in the superconducting state at
$T=0.06$\thinspace K and $\mu_{0}H=$0.5\thinspace T with a total counting time
of 52\thinspace hours in order to obtain good statistics and a small error
bar. The difference between $\chi(n)$ and the $\chi$($T$=0.06~K, $\mu_{0}%
H$=0.5\thinspace T) point demonstrates a clear drop in $\chi$ of $34\pm
6$\thinspace\% on entering the superconducting state.

Fig.~\ref{fig:Figure}(b) shows PNS and NMR results presented as a function of
magnetic field. It is notable that PNS yields a larger susceptibility in the
superconducting state than what can be deduced from the $^{17}$O Knight shift
\cite{Ishida2019_IsMM} $K$ at comparable field and temperature. Specifically,
at $\mu_{0}H\simeq0.5\,T$ and $T\simeq60$\thinspace mT, we have $\chi
(\text{PNS})/\chi(n)=0.66\pm0.06$ and $K(\text{NMR})/K(n)=0.37\pm0.08$. In the
Supplementary Material \cite{Supp_mat}, we show how the measured
susceptibility $\chi$ is corrected for Fermi liquid effects (Stoner
enhancement) and the orbital contribution. The corrected non-interacting spin
susceptibility $\chi_{0}$, which can be compared with theory, is shown in
Fig.~\ref{fig:Figure}(c).

As discussed above, PNS and NMR probe the magnetization in different ways so
there are reasons why the results may be different. For the (011) Bragg
reflection, the PNS method measures $M[\mathbf{G}=(011)]\approx M_{\text{Ru}%
}\times f_{Ru}(\mathbf{G})+1.04\times M_{O(2)}\times f_{O}(\mathbf{G})\approx
M_{\text{Ru}}\times f_{Ru}(\mathbf{G})$, where $f$ is the magnetic form factor
and $M$ is the magnetic moment on a given site. Note that the O(1) oxygen
sites [see Fig.~\ref{fig:Figure}(c)] do not contribute to the magnetic signal
observed at this reflection. Further, the moment on the oxygen O(2) sites is
known to be small from NMR Knight shift measurements~\cite{Ishida2019_IsMM}
and DFT calculations~\cite{Haverkort2008_HETS,Noce1999_NoCu}. Thus, our
measurement is essentially sensitive only to the Ru sites where most of the
moment resides. In contrast, recent NMR
experiments~\cite{Pustogow2019_PLCS,Ishida2019_IsMM} probed the oxygen O(1) sites.

As mentioned, $\chi_{0}(T)$ for an isotropic $s$-superconductor is expected to
follow the Yosida function \cite{Yosida1958_Yosi}. This function can be easily
modified~\cite{Won1994_WoMa} to account for a superconductor with vertical
line nodes. At low temperatures the drop in $\chi\equiv M/H$ will be
field-dependent due to the introduction of vortices by the magnetic field.
Modeling the effect of vortices on the spin susceptibility is theoretically
difficult; in addition, Sr$_{2}$RuO$_{4}$ shows a first order phase transition
at $H_{c2}$ with a step in the spin magnetization
\cite{Deguchi2002,Kittaka2014,Amano2015_AIIN}. Nevertheless, in
Fig.~\ref{fig:Figure}(c) we fit two simple illustrative low-temperature field
dependencies of $\chi_{0}(H)$ with $\chi_{0,s}(H)=\chi_{0}(0)+\Delta
\chi\,H/H_{c2}$ and $\chi_{0,\mathrm{nodal}}(H)=\chi_{0}(0)+\Delta\chi
\,\sqrt{H/H_{c2}}$ for the $s$-wave and nodal cases using the actual $H_{c2}$
\cite{Caroli1964a,Volovik1993a}. Our $s$-wave and nodal field-dependent fits
yield zero-field residual values of $\chi_{0}(0)/\chi_{0}(n)$ of $0.55\pm0.09$
and $0.29\pm0.15$ respectively. Both fits give $\chi\approx8\times10^{-4}%
$\thinspace$\mu_{B}$\thinspace T$^{-1}$\thinspace f.u.$^{-1}$ for $\mu_{0}%
H=1$\thinspace T.

Our data is consistent with two interesting scenarios as $H \rightarrow0$: (i)
a rapid reduction of $\chi_{0}$ below $\tfrac{1}{3}H_{c2}$ (solid line in
Fig.~\ref{fig:Figure}(c)); (ii) a large residual contribution to $\chi_{0}$ in
the $H \rightarrow0$ limit (dotted line). Case (i) would be consistent with an
even-parity (singlet) state with deep minima or nodes \cite{Volovik1993a} in
the gap and a small residual spin susceptibility. In case (ii) there would be
a large residual spin susceptibility. This would be qualitatively consistent
with various states in Table~\ref{tab:irr_reps} and other proposals discussed below.

Both $\chi_{0}(H)$ and the linear coefficient of specific heat $\gamma
(H)=C_{e}/T$ can detect the low-energy states introduced by vortices. For a
singlet superconductor, they are expected to show similar behavior
\cite{Amano2015_AIIN}. In Fig.~\ref{fig:Figure}(b,c) we reproduce the measured
\cite{Kittaka2018_KNSK} $\gamma(H)$ for $\mathbf{H}\Vert\lbrack100]$ and the
recent NMR Knight shift~\cite{Pustogow2019_PLCS,Ishida2019_IsMM} which both
yield smaller values when normalized to the normal state \cite{Supp_mat}.

It is instructive to compare the present PNS data with that of Duffy et
al.~\cite{Duffy2000_DHMM} measured with $\mathbf{H}\Vert\lbrack110]$. From
Table~\ref{tab:irr_reps} one can see that the effect of the field is expected
to be the same for the [100] and [110] directions for all order parameter
symmetries, except for two of the $E_{u}$ states. In Fig.~\ref{fig:Figure}(a),
we also show the PNS results of Duffy et al.~\cite{Duffy2000_DHMM} (open
squares) measured at a field of 1~T with $\mathbf{H}\Vert\lbrack110]$. These
were probed at the (002) Bragg peak and have therefore been scaled by the
ratio of the Ru form factors \cite{Brown1992_Brow,Ru_form} at (011) and (002)
for comparison. The solid and dotted lines show the expected $T$-dependence
based on the fitted $\chi(H)$ in Fig.~\ref{fig:Figure}(b,c) and the Yosida
function \cite{Yosida1958_Yosi,Won1994_WoMa}. It is likely that Duffy et
al.~\cite{Duffy2000_DHMM} were unable to resolve a change because of the lower
statistical accuracy and use of a higher field, where the suppression effect
is smaller due to the field induced density of states. However, the $E_{u}$ state
$\mathbf{d}=(\hat{\mathbf{x}}-\hat{\mathbf{y}})k_{z}$  (which shows no change
in $\chi_{0}$ for $\mathbf{H}\Vert\lbrack110]$ and a change for $\mathbf{H}%
\Vert\lbrack100]$) cannot currently be ruled out by the PNS experiments.

\textit{Discussion.}---Many superconducting states have been proposed for
Sr$_{2}$RuO$_{4}$. Some of those are shown in Table~\ref{tab:irr_reps} and,
following the observations of Pustogow et al.~\cite{Pustogow2019_PLCS}, there
have also been new theoretical proposals
\cite{Romer2019_RSEH,Roising2019_RSFL,Kivelson2020}. The PNS measurements
reported here yield a non-interacting spin susceptibility in the
superconducting state $\chi_{0}(\mu_{0}H=0.5~\text{T})/\chi_{0}(n)=0.71\pm
0.06$ which is larger than the NMR Knight shift
\cite{Pustogow2019_PLCS,Ishida2019_IsMM}. Thus the PNS data better match
different states. We concluded above that the residual $\chi_{0}(0)/\chi
_{0}(n)=0.55\pm0.09$ or $0.29\pm0.15$ for non-nodal [e.g. $\hat{\mathbf{x}%
}k_{x}+\hat{\mathbf{y}}k_{y}$, $\hat{\mathbf{z}}(k_{x}\pm ik_{y})$] or (near)
nodal [e.g. $s^{\prime}$, $d_{x^{2}-y^{2}}$, $(k_{x}\pm ik_{y})k_{z}]$ gaps
respectively. Thus, our PNS measurements do not rule out all odd-parity
states, but they do rule out those with $\chi_{0}(0)/\chi_{0}(n)=1$ in
Table~\ref{tab:irr_reps}. This includes the previously widely considered
chiral $p$-wave $\mathbf{d}=\hat{\mathbf{z}}(k_{x}\pm ik_{y})$ state
\cite{Rice1995_RiSi,Baskaran1996_Bask,Mackenzie2003_MaYo,Maeno2012}.
Odd-parity states with in-plane $\mathbf{d}$ vectors such as the helical
triplet (e.g. $\mathbf{d}=\hat{\mathbf{x}}k_{x}+\hat{\mathbf{y}}k_{y}$) states
proposed by R\o {}mer et al. \cite{Romer2019_RSEH} have a partial ($\approx
50$\%) suppression of $\chi_{s}$ and therefore are not ruled out. Other states
that are qualitatively compatible with our observations include the TRSB
$s^{\prime}+id_{x^{2}-y^{2}}$ and non-unitary $\hat{\mathbf{x}}k_{x}\pm
i\hat{\mathbf{y}}k_{y}$ states \cite{Romer2019_RSEH} proposed by R\o{}mer
\textit{et al}. \cite{Romer2019_RSEH}, states resulting from the 3D model of
R\o{}ising et al.~\cite{Roising2019_RSFL}, the TRSB $d_{xz\pm iyz}$ of Zutic and Mazin\cite{Zutic,Suh2019}, or the newest $d+ig$ proposal by Kivelson et al. \cite{Kivelson2020}.

The smaller $\chi_{0}(0)/\chi_{0}(n)=0.29\pm0.15$ for nodal states is
compatible with even-parity order parameters with deep minima or nodes in the
gap, listed above. Particularly, if other (impurity or orbital) contributions
are included in our estimated $\chi_{0}$ which are not due to quasiparticles
excitations \cite{Abrikosov1962AbGo,Bang2012,Supp_mat}. The nodal $s^{\prime}%
$-wave and $d_{x^{2}-y^{2}}$-wave states are supported by thermal conductivity
\cite{Hassinger2017_HBTR}, angle-dependent specific heat
\cite{Kittaka2018_KNSK}, penetration depth \cite{Bonalde2000_BYSV} and
quasiparticle interference experiments \cite{Sharma2019_SEWK}, albeit not by
the evidence of TRSB, or by the recently observed discontinuity in the
$c_{66}$ shear modulus\cite{Gosh2020}. Future polarized neutron scattering
measurements at lower fields and other field directions will place further
constraints on the allowed paired states.

The authors are grateful for helpful discussions with J.\;F.\;Annett,
S.\;E.\;Brown, K.\;Ishida, A. T.\;R\o{}mer, and S.\;Yonezawa. Work is supported by EPRSC Grant
EP/R011141/1, the Centre for Doctoral Training in Condensed Matter Physics
(EPRSC Grant EP/L015544/1), JSPS Kakenhi (Grants JP15H5852 and JP15K21717) and
the JSPS-EPSRC Core-to-Core Program ``Oxide-Superspin (OSS)''.

\vfill
\bibliographystyle{apsrev4-1}
\bibliography{Sr2RuO4}

\begin{thebibliography}{62}
\bibitem [{Ru_()}]{Ru_form}%
  \BibitemOpen
  \href@noop {} {}\bibinfo {note} {The available form factor {Ru$^{+}$} was
  used.}\BibitemShut {Stop}%
\end{thebibliography}

\vfill

\end{document}


\title{Supplementary Material:\\Orbital contributions to the susceptibility, Fermi liquid corrections and a
two fluid model}

\begin{abstract}
Our polarized neutron scattering measurement probes the total magnetic moment
(or associated susceptibility) including spin and orbital contributions. In
addition, the spin susceptibility in Sr$_{2}$RuO$_{4}$ is subject to an
exchange or Stoner enhancement. The effect of the Stoner enhancement is
temperature and field dependent. Here we describe how these two effects can be
taken into account when comparison is made with theory.

\end{abstract}
\maketitle

\section{Background}

Theoretical models describing the spin susceptibility in superconductors such
as Sr$_{2}$RuO$_{4}$ typically aim to calculate the non-interacting
susceptibility. In the normal state this would be $\mu_{B}^{2} \mathcal{N}%
_{F}$ and would be obtained from a density functional theory (DFT) band
structure calculation, where the non-interacting unrenormalized electron
density of states at the Fermi energy is $\mathcal{N}_{F}$. In Landau Fermi
Liquid theory \cite{Coleman2015}, the $H \rightarrow0 $ non-interacting
renormalized spin susceptibility $\chi_{0}$ and specific heat coefficient
$\gamma=C_{e}/T$ are given by $\chi_{0}=\mu_{B}^{2} \mathcal{N}^{\star}_{F}$
and $\gamma=(\pi^{2}/3)\; k_{B}^{2} \mathcal{N}^{\star}_{F}$, where
$\mathcal{N}^{\star}_{F}$ is the quasiparticle density of states.

Polarized neutron scattering measurements probe the total magnetic moment (or
associated susceptibility) including spin and orbital contributions. In
addition, $\chi_{0}$ in Sr$_{2}$RuO$_{4}$ is subject to a Stoner enhancement
where the Stoner factor is $I$. Our experiments are performed in a magnetic
field, where we define the susceptibility $\chi_{0}(H)=M/\mu_{0}H$ and $\chi_{0}(n)$
denotes the normal state value (say at $H_{c2}$).

The susceptibility due to the orbital and spin components is then given by
\begin{align}
\label{eqn:chi_total}
\chi(H)=\chi_{\mathrm{orb}}+\chi_{\mathrm{spin}}(H)=\chi_{\mathrm{orb}}%
+\frac{\chi_{0}(H)}{1-I\chi_{0}(H)}.
\end{align}
DFT calculations estimate $\chi_{\mathrm{orb}}\approx1.5\times10^{-4}\,\mu
_{B}\mathrm{T}^{-1}\mathrm{f.u.}^{-1}$, although, DMFT suggests an enhancement
by fluctuations of the orbital contribution up to a factor of
two~\cite{Kim2018_KMFP,Tamai2019_TZRC} to $\sim3\times$10$^{-4}$\thinspace
$\mu_{B}$T$^{-1}$f.u.$^{-1}$. It will be convenient to define the ratio of the
orbital and spin parts of the susceptibility in the normal state $f$ as
\begin{align}
f=\frac{\chi_{\mathrm{orb}}}{\chi_{\mathrm{spin}}(n)}=\chi_{\mathrm{orb}}%
\frac{1-I\chi_{0}(n)}{\chi_{0}(n)}.
\end{align}
Based on the DFT calculations, we estimate $f=0.11$.

We may use measured normal state values of the susceptibility and linear
coefficient of the electronic specific heat $\gamma$ to estimate Stoner enhancement. 
The measured susceptibility for fields parallel
to the $a$-axis is $\chi_{a}=0.91\,\mbox{emu mole}^{-1}=1.61\times10^{-3}%
\,\mu_{B}T^{-1}\mathrm{f.u.}^{-1}$\cite{Ishida1998_IMKA}. From the measured
susceptibility and the estimate of $\chi_{\mathrm{orb}}$ we obtain
$\chi_{\mathrm{spin}}(n)=1.46\times10^{-3}\,\mu_{B}T^{-1}\mathrm{f.u.}%
^{-1}=1.04\times10^{-8}\,\mbox{mole}^{-1}$. The linear specific heat is linear
specific heat $\gamma(n)=40\,\mbox{mJ mole}^{-1}\mbox{K}^{-2}$, yielding a
Wilson ratio $W=(\chi/\gamma)\times(\pi^{2}k_{B}^{2})/(3\mu_{0}\mu_{B}%
^{2})=1.5$. Assuming a Stoner enhancement factor equal to the Wilson ratio, the last term of
Eq.~\ref{eqn:chi_total} is
\[
W=\frac{1}{1-I\chi_{0}(n)}.
\]
We can estimate $I\chi_{0}(n)=S=0.33$. The enhancement of the spin
susceptibility can also be understood through Fermi liquid theory ($F_{0}^{a}$
parameter) \cite{Mazin1997_MC}. In the case of $^{3}$He, which has $W\approx3$, Fermi liquid effects
are required to understand the temperature dependence of the spin
susceptibility of the superfluid phases \cite{Leggett1965}.

\section{Correction for orbital susceptibility and Stoner enhancement}

In Fig.\thinspace1 of the main paper we plot susceptibilities normalized to
their normal state values as a function of field and temperature. We should
remember that in the case of the polarized neutron scattering we measure
$M(\mathbf{G})/\mu_{0}H$ in absolute units. This susceptibility contains a
form factor due to the electron orbital hence the values are lower those
quoted above by a factor of 0.68\cite{Brown1992_Brow}. We define relative
susceptibilities
\[
X(H)=\chi(H)/\chi(n),
\]
and
\[
X_{0}(H)=\chi_{0}(H)/\chi_{0}(n),
\]
where $\chi(H)$ refers to the measured susceptibility in
Eq.~\ref{eqn:chi_total} and $\chi_{0}(H)$ refers to the corresponding
non-interacting spin susceptibility. Fig.\thinspace1(a,b) shows the measured
$X(B,T)$. $X_{0}(H)$ and $X(H)$ are related through the equations
\begin{align}
X(H) &  =\frac{\displaystyle\chi_{\mathrm{orb}}+\frac{\chi_{0}(H)}{1-I\chi
_{0}(H)}}{\displaystyle\chi_{\mathrm{orb}}+\frac{\chi_{0}(n)}{1-I\chi_{0}(n)}%
}\\
&  =\left(  X_{0}(H)+\frac{f[1-X_{0}(H)]}{(f+1)(1-S)}\right)  \frac
{1-S}{1-SX_{0}(H)}\\
X_{0}(H) &  =\frac{f(X(H)-1)+X(H)}{1+(1+f)S(X-1)}
\label{eqn:X_X0}%
\end{align}
Thus we use Eqn.~\ref{eqn:X_X0} with $f=0.11$ and $S=0.33$ to correct for
orbital susceptibility and Stoner enhancement of the susceptibility and relate
the data and curves in Fig.\thinspace1(b) and (c). In particular, as discussed
in the main text, theory makes predictions for the field dependencies of the
non-interacting spin susceptibility $\chi_{0,s}(H)=\chi_{0}(0)+\Delta\chi
_{0}H/H_{c2}$ and $\chi_{0,\text{nodal}}(H)=\chi_{0}(0)+\Delta\chi_{0}%
\sqrt{H/H_{c2}}$.

\begin{table}[h]
\begin{center}
\begin{tabular}{|@{\;}c@{\;}|@{\;}c@{\;}|@{\;}c@{\;}|}
\hline
$X(H)=\chi(H)/\chi(n)$ & $X(H)$ & $X_{0}(H)$\\ \hline
$\chi(\mu_{0} H=0.5 \mbox{T})$ & 0.66(6) & 0.71(6)\\
$\chi(0), \chi_{0}(0)$ ($s$-wave) & 0.47(9) & 0.55(9)\\
$\chi(0), \chi_{0}(0)$ (nodal) & 0.16(14) & 0.29(15)\\ \hline
\end{tabular}
\end{center}
\caption{Measured and corrected susceptibilities. The table shows the effect
of the correction for the orbital susceptibility and Stoner enhancement (Fermi
liquid effects) on the susceptibility in the superconducting state.}%
\end{table}

\section{Two-fluid model for spin susceptibility and heat capacity}

As mentioned above, the linear coefficient of the electronic specific heat
$\gamma=C_{e}/T$ and the non-interacting spin susceptibility $\chi_{0}$ should
be proportional to the quasiparticle density of states $\mathcal{N}^{\star}_{F}$. This would imply that
$\chi_{0}$ and $\gamma$ are also proportional to each other and hence,
the relative values $\gamma(H)/\gamma(n)$ and $X_{0}(H)=\chi_{0}(H)/\chi
_{0}(n)$ should be equal. In Fig.\thinspace1(c) of the main paper the
relative values $\gamma(H)/\gamma(n)$ and $\chi_{0}(H)/\chi
_{0}(n)$ are shown, however, they display different field dependencies. 
This could be due to various causes (i-iv). (i) It could be that the orbital susceptibility in
Sr$_{2}$RuO$_{4}$ is larger than the applied value of $1.5\times10^{-4}%
\,\mu_{B}\mathrm{T}^{-1}\mathrm{f.u.}^{-1}$ calculated by DFT and is instead
closer to the fluctuation enhanced doubled value predicted by
DMFT~\cite{Kim2018_KMFP,Tamai2019_TZRC}. (ii) The sample we probed by PNS has
a larger residual $\gamma$ than the sample used in the specific heat
measurements reproduced in Fig.\thinspace1(b,c) in the main
paper\cite{Kittaka2018_KNSK}. Sample dependence of the linear coefficient
$\gamma$ has been reported in Sr$_{2}$RuO$_{4}$ in the
past\cite{Kittaka2018_KNSK,NishiZaki2000_NiMM}. (iii) There is a nuclear
polarization of some of the Sr and Ru nuclei due to the applied field and low
temperature, which would give an additional term in the PNS flipping ratio
\cite{Shull1963_ShFe,Ito1969_ItSh}. (iv) The two-fluid model, which describes
a supercondcutor as a superposition of a superconducting condensate and a
normal Fermi-liquid with no interference, is not a sufficiently accurate
description for the superconducting state in Sr$_{2}$RuO$_{4}$.

The measured $^{17}$O NMR Knight shifts reproduced in Fig.\thinspace
1(b,c)\cite{Pustogow2019_PLCS,Ishida2019_IsMM} in the main paper are as well
as the spin susceptibility subject to Stoner enhancement and should also
contain orbital contributions. If we assume the same Stoner factor (S=0.33),
that we observe that the relative non-interacting Knight shifts $K_{0}
(H)/K_{0}(n)$ analogue to $X_{0}$ in Eq.~\ref{eqn:X_X0} agrees with the
reproduced relative linear coefficient of the electronic specific heat
$\gamma(H)/\gamma(n)$ without any orbital contribution ($f=0$). 
\bibliographystyle{apsrev4-1}
\bibliography{Sr2RuO4}

\includegraphics[width=0.9\linewidth]{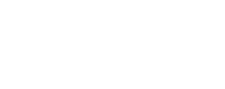}